\documentclass{ws-p8-50x6-00}

\begin{document}

\title{Experimental status of pionium at CERN}


\author{J. Schacher\\
on behalf of the DIRAC Collaboration}

\address{Universit\"at Bern, CH-3012 Bern, Switzerland}


\maketitle


The experiment DIRAC~\cite{di} aims to measure 
the lifetime of the $\pi^+\pi^-$ atom (pionium) 
in the ground state with $10\%$ precision. 
DIRAC is a magnetic double arm spectrometer and 
uses the high intensity 24~GeV/c 
proton beam of the CERN PS. 
Since this lifetime $\tau$ of order $10^{-15}$~s is dictated by 
{\em strong} interaction at low energy, a precise measurement of 
$\tau$ enables to determine {\em strong} S-wave pion scattering lengths 
to $5\%$~\cite{ru}. In Chiral Perturbation Theory (ChPT) pion scattering 
lengths are studied since many years, resulting in better and better 
predictions, now at an accuracy around $2\%$~\cite{co}. Such 
a measurement would submit the predictions of ChPT and hence the
understanding of chiral symmetry breaking of QCD to a crucial test. 

The experimental method to measure the pionium lifetime ($\sim 3$~fs)
takes advantage of the Lorentz boost of relativistic pionia produced in
high energy proton nucleus reactions at 24 GeV/c. After production 
in the target (e.g. 0.1~mm thin Ni foil) these relativistic 
($\gamma\simeq 15$) atoms may either decay into 
2~$\pi^0$ or get excited or ionized in the target material. 
In the case of ionization or breakup, characteristic charged 
pion pairs, called ``atomic pairs'', will emerge, exhibiting 
low relative momentum in their pair system ($Q < 3$~MeV/c), 
small pair opening angle ($\theta_{+-}< 3$~mrad) and 
nearly identical energies in the lab system 
($E_+\simeq E_-$ at the 0.3\% level).
The task of the spectrometer is  
to identify charged pion pairs and to measure their relative 
momenta $Q$ with high resolution of $\delta Q \simeq 1$~MeV/c.
By these means, it is possible to determine the number of ``atomic pairs''
above background, arising from pion pairs in a free state (``free pairs''). 
For a given target material and thickness, the ratio of the number of observed
``atomic pairs'' to the total amount of produced pionia depends on
$\tau$ in a unique way.

The experiment consists of coordinate detectors, 
a spectrometer magnet (bending power of 2.3 Tm) and two
telescope arms, each equipped with drift chambers, scintillation
hodoscopes, gas Cherenkov counters, preshower and muon detectors. 

After tuning the primary proton beam as well as all detectors, 
the experiment started to accumulate data in 1999. 
The setup performance has been  
studied by analysing data collected with a platinum target. 
In order to track the relative momentum resolution, the 
invariant mass spectrum of p and $\pi^-$ 
--- $\Lambda$ decay products --- was investigated. 
The position of the mass peak in Fig.~1 
corresponds to $\Lambda$ particles 
reconstructed in the DIRAC spectrometer. 
The width of the $\Lambda$ peak, mainly given by 
the momentum resolution, is found to be 
$0.43$~MeV.

A preliminary search for ``atomic pairs'' in the 1999 platinum 
data sample is presented in Fig.~2.   
As mentioned above, ``atomic pairs'' are characterized by low relative momentum 
$Q < 3$ MeV/c or equivalently $F < 3$ 
\footnote{Definition:  
$F = \sqrt{(Q_L/{\sigma_{Q_L}})^2 +
           (Q_X/{\sigma_{Q_X}})^2 +
           (Q_Y/{\sigma_{Q_Y}})^2}$ with resolutions $\sigma_{Q}$.}. 
The distribution of ``reals'' $N_{\rm real}(F)$ is the sum of two distributions: 
``atomic pairs'' (for $F < 4$) and ``free pairs''. 
In the region $F > 4$ the ``free~pair'' distribution $N(F)$ has been  
approximated by a fit function $A(F)$, based on ``accidental pairs''. 
In Fig.~2 the difference between the experimental distribution 
$N_{\rm real}(F)$ and $A(F)$ is shown. The excess of $\pi^+\pi^-$ pairs 
with $F < 2$ amounts to $160 \pm 45$, which is compatible 
with the expected number of ``atomic pairs'' $\approx 240$.

\begin{figure}[h] 
\begin{center}

\begin{minipage}[t]{5.5cm}
\mbox{\epsfig{figure=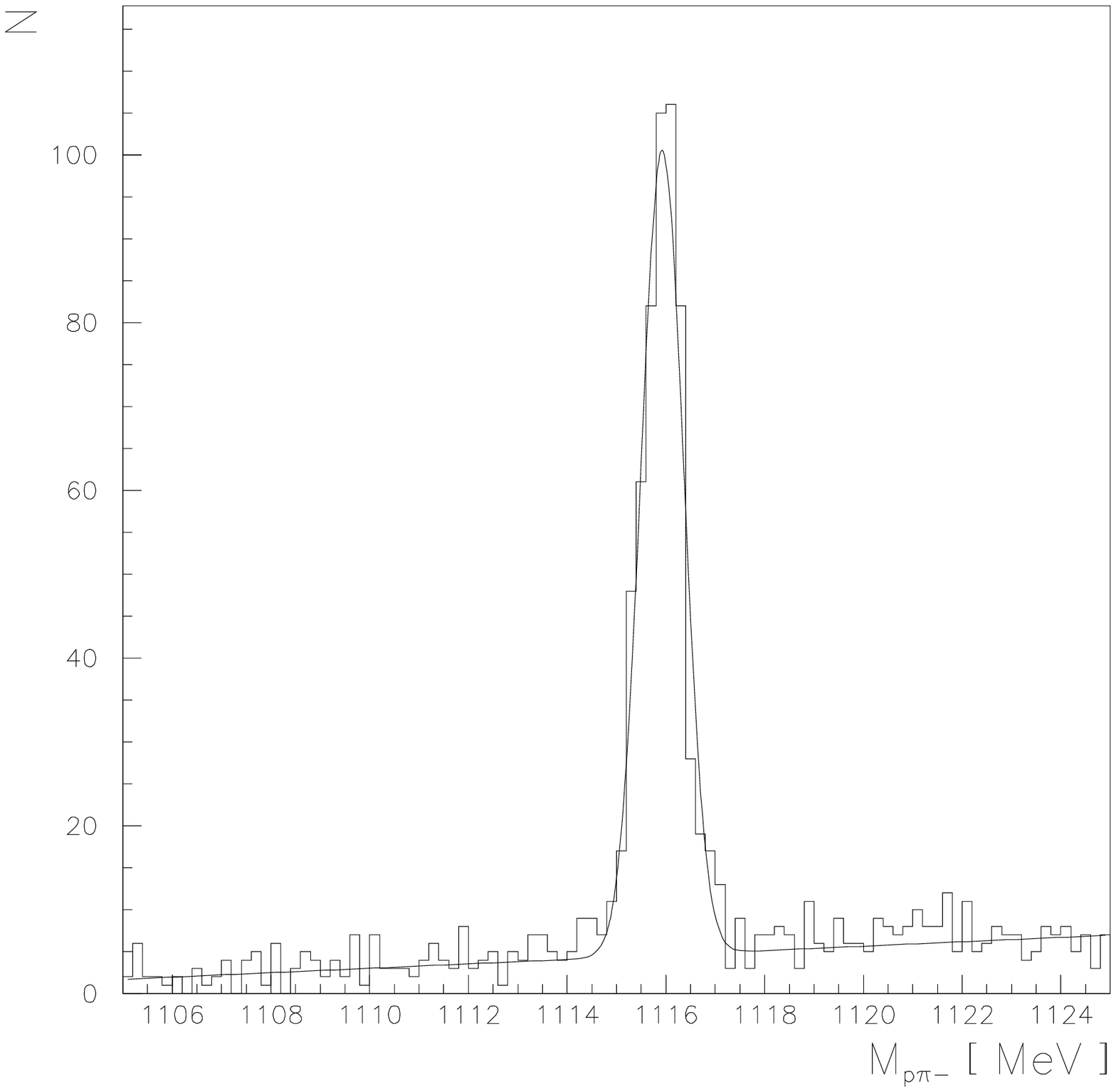,width=5.5cm}}
\caption{Invariant mass distribution of $p \pi^-$ 
         from $\Lambda$ decay ($\sigma_{\Lambda}=0.43$~MeV).} 

\end{minipage}
\hspace{0.5cm}
\begin{minipage}[t]{5.5cm}
\mbox{\epsfig{figure=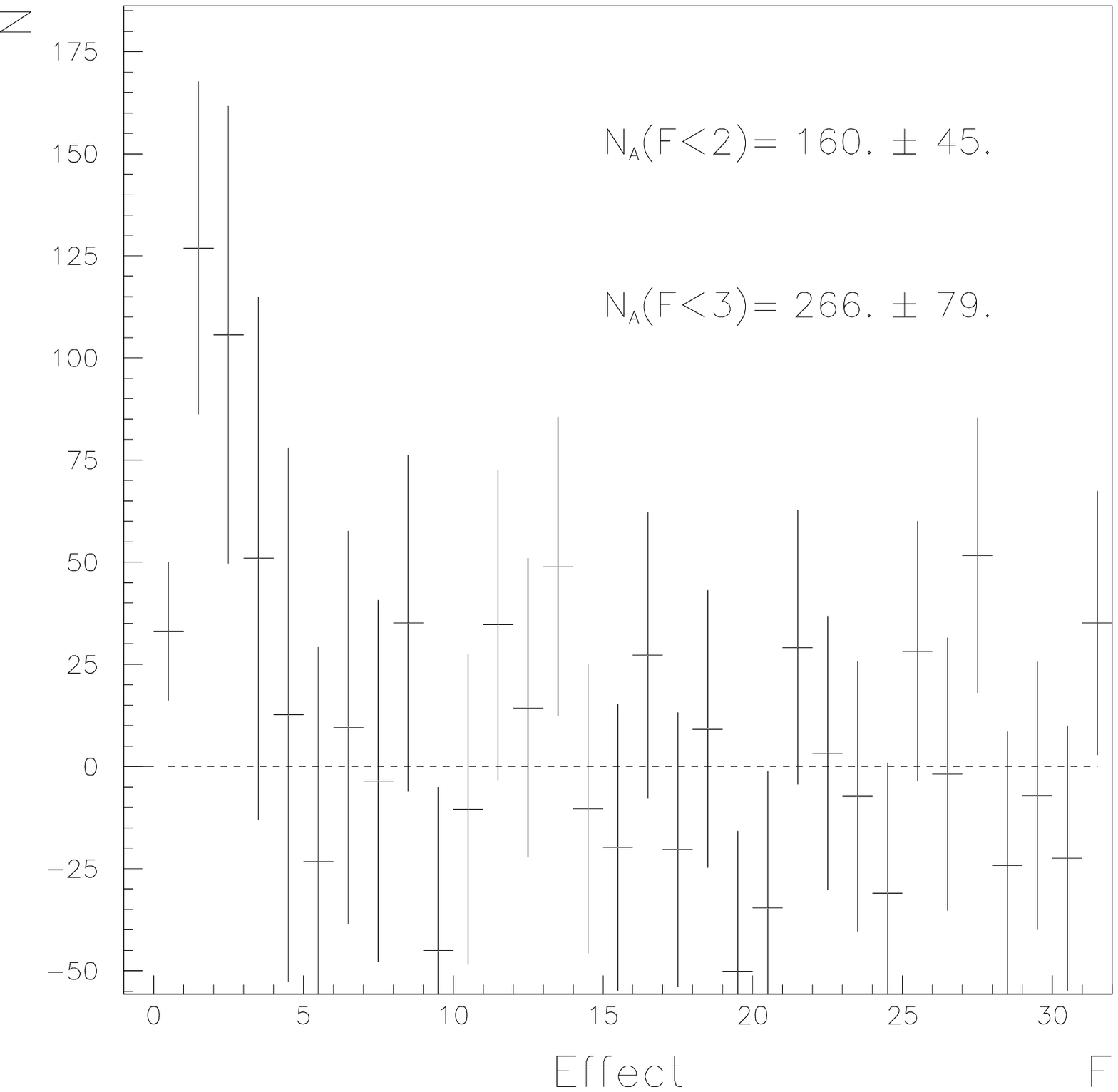,width=5.5cm}}
\caption{Candidates of ``atomic pairs'' in the $low~F$ region.}
\end{minipage}

\end{center}
\end{figure}

\end{document}